\documentclass[lettersize,journal]{IEEEtran}
\usepackage{amsmath,amsfonts}
\usepackage{algorithmic}
\usepackage{algorithm}
\usepackage{array}
\usepackage[caption=false,font=normalsize,labelfont=sf,textfont=sf]{subfig}
\usepackage{enumitem}
\usepackage{textcomp}
\usepackage{stfloats}
\usepackage{url}
\usepackage{verbatim}
\usepackage{graphicx}
\usepackage{cite}
\usepackage[hidelinks]{hyperref}
\hypersetup{
    colorlinks=true,
    citecolor=blue,   
    linkcolor=blue,   
    urlcolor=blue
}
\usepackage{footmisc}

\usepackage{booktabs}
\usepackage{multirow}
\usepackage{graphicx}
\usepackage{amssymb}
\usepackage{pifont}
\newcommand{\cmark}{\ding{51}}
\newcommand{\xmark}{\ding{55}}
\usepackage{threeparttable}
\usepackage{hyperref}

\begin{document}

\title{Task-disentangled Low-Rank Adaptation for Versatile Audio-visual Multi-modal Learning Tasks within a Unified  Framework}

\author{
Hanyu Xuan, 
Mengqi Zhang,
Junjun Mao,
Fei Wang,
Kun Li,
Guanghui  Yue,~\IEEEmembership{Senior Member,~IEEE},\\
Zhiliang Wu,~\IEEEmembership{Member,~IEEE},
Hehe Fan,~\IEEEmembership{Senior Member,~IEEE}
\thanks{This work was supported in part by the Natural Science Foundation of Anhui Province (No.2308085QF221), 
in part by the National Natural Science Foundation of China (No.62302006 and No.12271002) and the Earth System Big Data Platform of the School of Earth Sciences, Zhejiang University. (\emph{Corresponding author: Junjun Mao and Zhiliang Wu})}
\thanks{Hanyu Xuan, Mengqi Zhang, and Junjun Mao are with the School of Big Data and Statistics,
Anhui University, Hefei 230039, China.}
% 22176@ahu.edu.cn
% we24301059@stu.ahu.edu.cn
% maojunjun@ahu.edu.cn
\thanks{Fei Wang is with the Institute of Artificial Intelligence, Hefei Comprehensive National Science Center, Hefei, 230026, China. }
%jiafei127@gmail.com
\thanks{Kun Li is with the College of Information Technology, United Arab Emirates University, Abu Dhabi, 15551, United Arab Emirates.}
%kunli.hfut@gmail.com
\thanks{Guanghui  Yue is with the School of Biomedical Engineering, Shenzhen University, Shenzhen 518060, China.}
%yueguanghui@szu.edu.cn
\thanks{Zhiliang Wu is with the College of Computing and Data Science, Nanyang Technological University, Singapore 639798, Singapore.}
%zhiliang.wu@ntu.edu.sg
\thanks{Hehe Fan is with the School of Artificial Intelligence, Zhejiang University, Hangzhou 310007, China.}
%hehefan@zju.edu.cn
}

\markboth{IEEE Transactions on Multimedia,~Vol.~XX, No.~XX, Month~2026}%
{Shell \MakeLowercase{\textit{et al.}}: A Sample Article Using IEEEtran.cls for IEEE Journals}

\IEEEpubid{}
\maketitle

\begin{abstract}
Inspired by human multi-modal perception,
Audio-Visual Multi-Modal Learning (AVMML) 
integrates auditory and visual information 
to leverage complementary  cross-modal cues, 
enabling more robust and comprehensive scene perception.
Existing studies predominantly tackle each AVMML task in isolation,
which stands in stark contrast to humans' unified cognitive capacity for handling versatile perception.
However, naive joint training across multiple AVMML tasks often suffers from mutual interference, arising from the intricate inter-task relationships.
To address this, 
we propose a unified framework that simultaneously accommodates versatile AVMML tasks.
Specifically,
benefiting from powerful representation and generalization capabilities of large language models,
we design a task-disentangled  Low-Rank Adaptation (LoRA) mechanism 
that enables dynamic integration of both task-specific and task-shared knowledge, 
thereby facilitating effective multi-task collaboration.
The proposed task-disentangled LoRA comprises three components: a task-general low-rank matrix, task-specific modulation matrices, and cross-task collaboration experts,
which respectively capture universal audio-visual knowledge, 
decouple task-specific pattern, 
and exploit inherent inter-task correlations.
By unifying explicit collaboration 
from both model and task perspectives,
our approach not only surpasses existing unified audio-visual models across  multiple AVMML tasks, 
but also outperforms most task-specific models on certain AVMML tasks.
\end{abstract}


\section{Introduction}

The human capacity to perceive the world through multiple simultaneous  senses is termed multi-modal perception in cognitive science \cite{treichler1967training}. 
Among these modalities,  
vision and hearing serve as the dominant sensory channels,
conveying most environmental information and offering complementary cues that support scene comprehension \cite{wen2009multisensory,GUO202513}. 
Inspired by this  human capacity, 
the integration of auditory and visual information in intelligent systems has been demonstrated to yield substantial benefits, 
since audio-visual modalities provide abundant mutually reinforcing cues that jointly facilitate more robust and comprehensive perception.
For this purpose, Audio-Visual Multi-Modal Learning (AVMML) has emerged as a prominent and rapidly growing research field \cite{wei2022audioVisualReview}.

\begin{figure}
\vspace{-0.5cm}
  \centering
  \includegraphics[width=1.0\linewidth]{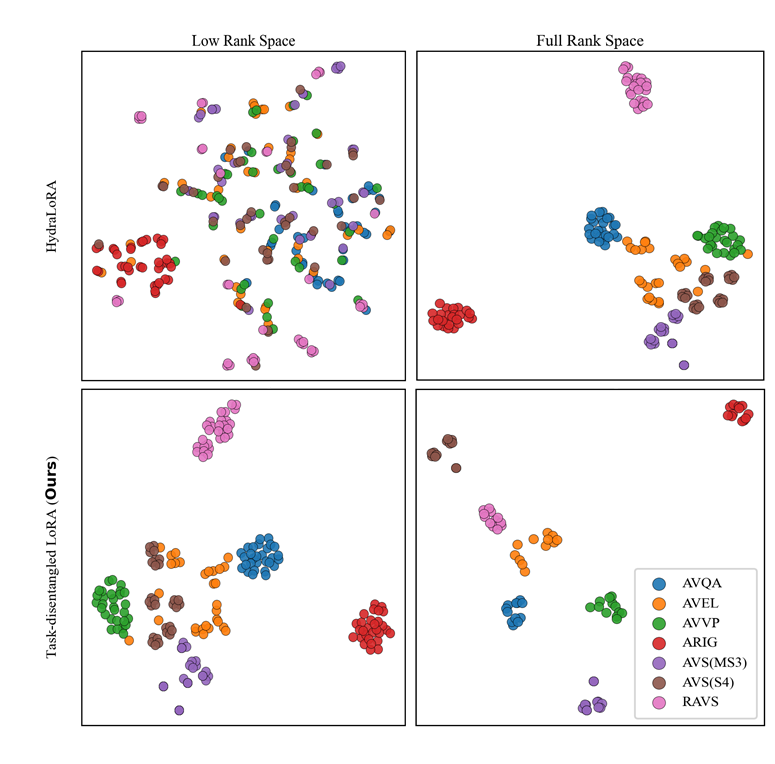}
\vspace{-1cm}
  \caption{T-SNE visualization of task-specific features extracted from the output linear layer of the final block in LLaMA-2-7B \cite{touvron2023llama}, comparing HydraLoRA \cite{tian2024hydralora} and the proposed LoRA after fine-tuning on six AVMML tasks.}
  \label{fig:T-sne}
\end{figure}

AVMML encompasses a diverse spectrum of core tasks, which can be primarily categorized into temporal localization, spatial localization,  pixel-level understanding, and spatio-temporal reasoning \cite{wei2022audioVisualReview,li2026av}.
Temporal localization focuses on predicting event occurrences within video sequences  and delineating their precise temporal boundaries,
with representative tasks including 
Audio-Visual Event Localization (AVEL) \cite{tian2018audio,linWang2019,lin2020audiovisual,xu2020cross,Owens_2018_ECCV,Zhou_2021_CVPR,yu2022mm}, 
Audio-Visual Video Parsing (AVVP) \cite{rachavarapu2024prototype,lamba2021crossmodal,chen2024cm,pasi2022modalityBias,wu2022perceptive,tian2020unified,wu2021heterogeneous,mo2022mmgn}.
In contrast,
spatial localization aims at pinpointing the positions of sounding objects in visual scenes,
as exemplified by 
Audio-Referred Image Grounding (ARIG) \cite{chen2021localizing,mo2022localizing,song2022self,liu2022exploiting,mo2022SLAVC,park2023marginnce,fedorishin2022hear,sun2023learning}.
Pixel-level understanding seeks to generate fine-grained segmentation masks for target objects, 
prominent tasks include Audio-Visual Segmentation (AVS) \cite{qian2020multiple,duke2021sstvos,Mao2021TransformerTS,zhang2021learning,zhou2022audio,liu2024bavs} and Reference AVS (RAVS) \cite{zhou2022audio,li2023robust,wang2024prompting,wu2022language,gao2024avsegformer,wang2024ref}.
Spatio-temporal reasoning jointly models temporal dynamics and spatial layouts to support high-level inference, 
with a typical task being Audio-Visual Question Answering (AVQA) \cite{2020Temporal,2015VQA,fan2019heterogeneous,yun2021panoavqa,lao2023coca,li2023progressive,lin2023vision,li2024object}.

While substantial  progress has been achieved 
on these AVMML tasks, 
existing  methods   
have primarily addressed each task in isolation.
In contrast, 
the human perceptual and cognitive system inherently  possesses a unified, multi-task perceptual and cognitive capacity for comprehending complex audio-visual scenes \cite{calvert1999response,ZHENG202524}.
To bridge this gap, 
Du et al. \cite{du2025crab} introduced
a unified framework
capable of handling a diverse range of  AVMML tasks.
Specifically,
this framework achieves parameter-efficient adaptation mechanism via a HydraLoRA \cite{tian2024hydralora} architecture,
which consists of a shared low-rank matrix and multiple independent LoRA heads,
whose outputs are dynamically aggregated via latent routing scores.

Despite this advancement,
this design suffers from two key limitations. 
First, audio-visual inputs from all AVMML tasks are indiscriminately mapped into the same low-dimensional subspace,
which degrades the discriminability of task-specific features \cite{yang2026disentangling}.
Second, 
the routing mechanism relies exclusively on audio-visual input without explicit task identity constraints, 
resulting in highly task-coupled representations.
Consequently,
this induces pronounced inter-task interference among AVMML tasks with divergent optimization objectives and heterogeneous feature distributions \cite{liu2021conflict}.
Empirical feature visualization in Fig.\ref{fig:T-sne} 
provides  further corroborating evidence for  the above analysis.
Under HydraLoRA-based adaptation,
features from  distinct AVMML tasks exhibit pronounced entanglement, particularly within the low-rank subspace.
This observation suggests that without explicit task disentanglement, input-dependent routing alone is insufficient to maintain well-separated task-specific feature distributions.

To tackle the above limitations, 
we propose a unified framework that simultaneously accommodates versatile AVMML tasks.
Specifically,
benefiting from powerful representation and generalization capabilities of Large Language Models (LLM),
we design  a task-disentangled Low-Rank Adaptation (LoRA) architecture and integrate it into the multi-modal LLMs.
This enables the dynamic adaptation and collaboration  of both task-shared  and task-specific knowledge across  diverse AVMML tasks.
The proposed task-disentangled LoRA first projects the input into a shared low-rank subspace to capture universal audio-visual knowledge.
To enhance task discriminability and preserve task-unique information,
we introduce multiple task-specific modulation matrices
to decouple task-specific patterns 
from this universal subspace.
Furthermore,
inspired by the Mixture-of-Experts (MoE) mechanism \cite{pmlr-v162-du22c},
we adopt multiple LoRA expert heads to facilitate adaptive and flexible cross-task knowledge transfer. 
Their outputs are dynamically aggregated via a task-adaptive routing, 
thereby facilitating effective collaboration  among various AVMML tasks.

% After task-guided adaptation is introduced, samples from the same task form more compact clusters, 
% while different tasks become more separable in both the low-rank and full-rank spaces
% We take this as evidence that task identity should not merely be implied by instruction prompts; 
% it should explicitly guide the construction of low-rank adaptation paths \cite{wang2024task,xu2025meteora,yang2026disentangling}.

As illustrated in Fig.\ref{fig:T-sne},
our method produces more compact
intra-task feature distributions and higher inter-task separability compared to   HydraLoRA \cite{tian2024hydralora},
in both low-rank and full-rank feature spaces.
To further evaluate task-level performance,
we evaluate the proposed method on six representative AVMML tasks,
covering  temporal localization, spatial localization,  pixel-level understanding, and spatio-temporal reasoning.
Comprehensive quantitative and qualitative results and analyses corroborate the effectiveness of the proposed method.
Our contributions can be summarized as follows:
\begin{itemize}
    \item We propose  a unified framework that accommodates versatile audio-visual multi-modal learning tasks.
    \item We design a task-disentangled  LoRA that enables dynamic integration of both task-specific and task-shared knowledge, facilitating effective multi-task collaboration.  
    \item Extensive experiments conducted on six representative tasks validate the superior performance, strong generalization ability, and universal effectiveness of our method.
\end{itemize}

% TG-MTL-LoRA on a unified audio-visual framework built on LLaMA-2-7B\cite{touvron2023llama}, spanning seven representative tasks that include temporal localization, spatial localization, spatio-temporal reasoning, and pixel-level understanding. 
% Compared with the original HydraLoRA-based adaptation strategy, TG-MTL-LoRA improves most metrics, with particularly clear gains on temporal localization, spatial localization, and pixel-level segmentation. 

% Ablation results confirm the necessity of both the task-specific modulation matrix and multiple expert heads, and the expert weight analysis reveals that different tasks learn distinct expert preferences. Taken together, these findings support the effectiveness of task-guided adaptation for unified audio-visual learning.

% \section{Related Work}
\section{Related Work}

AVMML encompasses a suite of core tasks, which can be primarily categorized into four types: temporal localization, spatial localization, pixel-level understanding, and spatio-temporal reasoning~\cite{wei2022audioVisualReview,li2026av}.

% \noindent 
\textbf{1) Temporal Localization}
focuses on predicting event occurrences in videos and delineating their corresponding temporal boundaries~\cite{li2026esg}.
It primarily comprises two representative tasks: AVEL and AVVP.
For AVEL, 
Tian et al.~\cite{tian2018audio} first formalize AVEL under both fully-supervised and weakly-supervised settings. 
Owens et al.~\cite{Owens_2018_ECCV} learn self-supervised multi-modal representations for audio-visual scene analysis. 
Lin et al.~\cite{linWang2019} introduce a dual-modality seq-to-seq framework to capture  temporal dependencies across audio and visual modality. 
Lin et al.~\cite{lin2020audiovisual} adopt a Transformer with instance attention to enable segment-level audio-visual interaction. 
% Subsequent works further enhance cross-modal interaction and event-relevant feature selection. 
% Specifically,
Xu et al.~\cite{xu2020cross} model fine-grained cross-modal relational dependencies.
Zhou et al.~\cite{Zhou_2021_CVPR} propagate positive samples along the audio-visual event line. 
Yu et al.~\cite{yu2022mm} exploit multi-scale temporal cues to boost event localization performance.

AVVP further requires the model to distinguish audible-only  events, visible-only events, and their co-occurring events over time. 
Tian et al.~\cite{tian2020unified} introduce the weakly-supervised AVVP task. 
% and learn heterogeneous representations for audio-visual events. 
% Subsequent methods improve AVVP by enhancing modality-specific learning and cross-modal correspondence modeling. 
% For instance,
Lamba et al.~\cite{lamba2021crossmodal} exploit cross-modal learning to achieve better event parsing. 
Wu et al.~\cite{wu2021heterogeneous} examine heterogeneous clues from different modalities. 
% In addition,
Pasi et al.~\cite{pasi2022modalityBias} introduce a modality-aware approach to analyze and alleviate modality bias in AVVP. 
Mo et al.~\cite{mo2022mmgn} design  a multi-modal grouping network that groups modality-specific temporal features.
Wu et al.~\cite{wu2022perceptive} develop a perceptive pretraining framework to enhance modality-specific event representations. 
% and improve audio-visual correspondence learning.
Chen et al.~\cite{chen2024cm} strengthen audio-visual interaction through cross-modal perception and interactive feature enhancement. 
Rachavarapu et al.~\cite{rachavarapu2024prototype} employ pseudo-labeling to generate more reliable segment-level supervision from video-level labels.
Chen et al.~\cite{chen2026teacher} introduce teacher-guided pseudo supervision and cross-modal alignment to improve weakly supervised audio-visual video parsing.
% Crab~\cite{du2025crab} incorporates both AVE and AVVP into a unified audio-visual scene understanding framework.

\textbf{2) Spatial Localization}
aims at grounding the positions of sounding
objects in visual scenes~\cite{sun2026rassu}.
% Most existing studies focus on audio-visual sound source localization, which learns the correspondence between acoustic signals and sounding objects without requiring spatial annotations.
% For instance, 
Chen et al.~\cite{chen2021localizing} discover audio-visual correspondences from unconstrained videos. 
Mo et al.~\cite{mo2022localizing} introduce a simple yet effective framework for visual sound localization. 
Song et al.~\cite{song2022self} replace contrastive negative sampling with a negative-free predictive learning objective.
Subsequent  works improve localization robustness.
For instance,
Liu et al.~\cite{liu2022exploiting} exploit transformation invariance and equivariance to learn more consistent localization representations.
Mo et al.~\cite{mo2022SLAVC} revisit weakly-supervised audio-visual localization and strengthen cross-modal correspondence learning.
% To mitigate the negative sample issue, 
Park et al.~\cite{park2023marginnce}  introduce a negative margin to reduce the adverse effects of semantically related negatives.
Fedorishin et al.~\cite{fedorishin2022hear} incorporate optical-flow-based motion cues to identify dynamic sounding objects.
Sun et al.~\cite{sun2023learning} develop a false-negative-aware contrastive objective to alleviate incorrect negative matching.
%Crab~\cite{du2025crab} further formulates audio-referred image grounding in a unified audio-visual scene understanding framework.

\textbf{3) Pixel-level Understanding}
seeks to generate fine-grained
segmentation masks for sounding or referred objects.
It primarily comprises two representative
tasks: AVS and RAVS.
For AVS,
Zhou et al.~\cite{zhou2022audio} formalize the AVS task and provide pixel-level annotations.
Qian et al.~\cite{qian2020multiple} progressively localize multiple sound sources via coarse-to-fine audio-visual correspondence learning.
Duke et al.~\cite{duke2021sstvos} adopt sparse spatio-temporal Transformers to propagate object masks across video frames.
Mao et al.~\cite{Mao2021TransformerTS} explore Transformer-based representations for salient and camouflaged object detection, which provide transferable visual knowledge for AVS.
Zhang et al.~\cite{zhang2021learning} integrate generative visual representations with an energy-based latent space to model uncertainty in saliency prediction.
Liu et al.~\cite{liu2024bavs} leverage knowledge from foundation models through a bootstrapping strategy to improve audio-guided mask prediction.
Recent studies further improve AVS by strengthening multimodal alignment and temporal representation learning.
Lv et al.~\cite{lv2026uncertainty} introduce uncertainty-aware dynamic fusion to mitigate unreliable cross-modal alignment during segmentation.
Shen et al.~\cite{shen2026av2ts} formulate AVS as a multivariate time-series modeling problem to capture temporal dependencies in audio-visual representations.
Zhuge et al.~\cite{zhuge2026temlo} jointly exploit temporal context and local visual details to enhance audio-visual segmentation.

RAVS further  extends AVS task by using language as referring cues to segment specific objects~\cite{zhou2026think}.
Li et al.~\cite{li2023robust} improve referring video object segmentation by enforcing structural consistency across video frames.
Wu et al.~\cite{wu2022language} encode language as queries to locate and track referred objects in videos.
Wang et al.~\cite{wang2024prompting} use acoustic signals as segmentation prompts to improve generalization across sounding objects and visual scenes.
Gao et al.~\cite{gao2024avsegformer} adopt Transformer-based cross-modal interaction to associate audio cues with pixel-level visual representations.
Wang et al.~\cite{wang2024ref} strengthen  the alignment  among audio, visual, and textual modalities for referred-object segmentation.

\textbf{4) Spatio-temporal Reasoning}
involves jointly modeling temporal dynamics and spatial layouts to support higher-level inference.
A representative task is AVQA.
Early efforts include Fayek et al.~\cite{2020Temporal}, who propose audio question answering as a task to evaluate temporal reasoning over acoustic events.
Fan et al.~\cite{fan2019heterogeneous} explore heterogeneous memory representations to store and retrieve multi-modal cues in dynamic videos.
Yun et al.~\cite{yun2021panoavqa} extend AVQA to panoramic videos, 
requiring grounded reasoning over broader visual contexts and spatial acoustic cues.
Recent studies further improve AVQA by strengthening causal reasoning, spatio-temporal perception, and multi-modal adaptation.
Lao et al.~\cite{lao2023coca} introduce collaborative causal regularization to mitigate spurious audio-visual correlations.
Li et al.~\cite{li2023progressive} progressively aggregate spatial and temporal evidence for question-oriented reasoning.
Lin et al.~\cite{lin2023vision} adapt pretrained vision Transformers for parameter-efficient audio-visual learning, 
while Li et al.~\cite{li2024object} enhance object-aware correspondence to improve evidence selection.

\section{Methodology}

\begin{figure}
  \centering
  \includegraphics[width=1.0\linewidth]{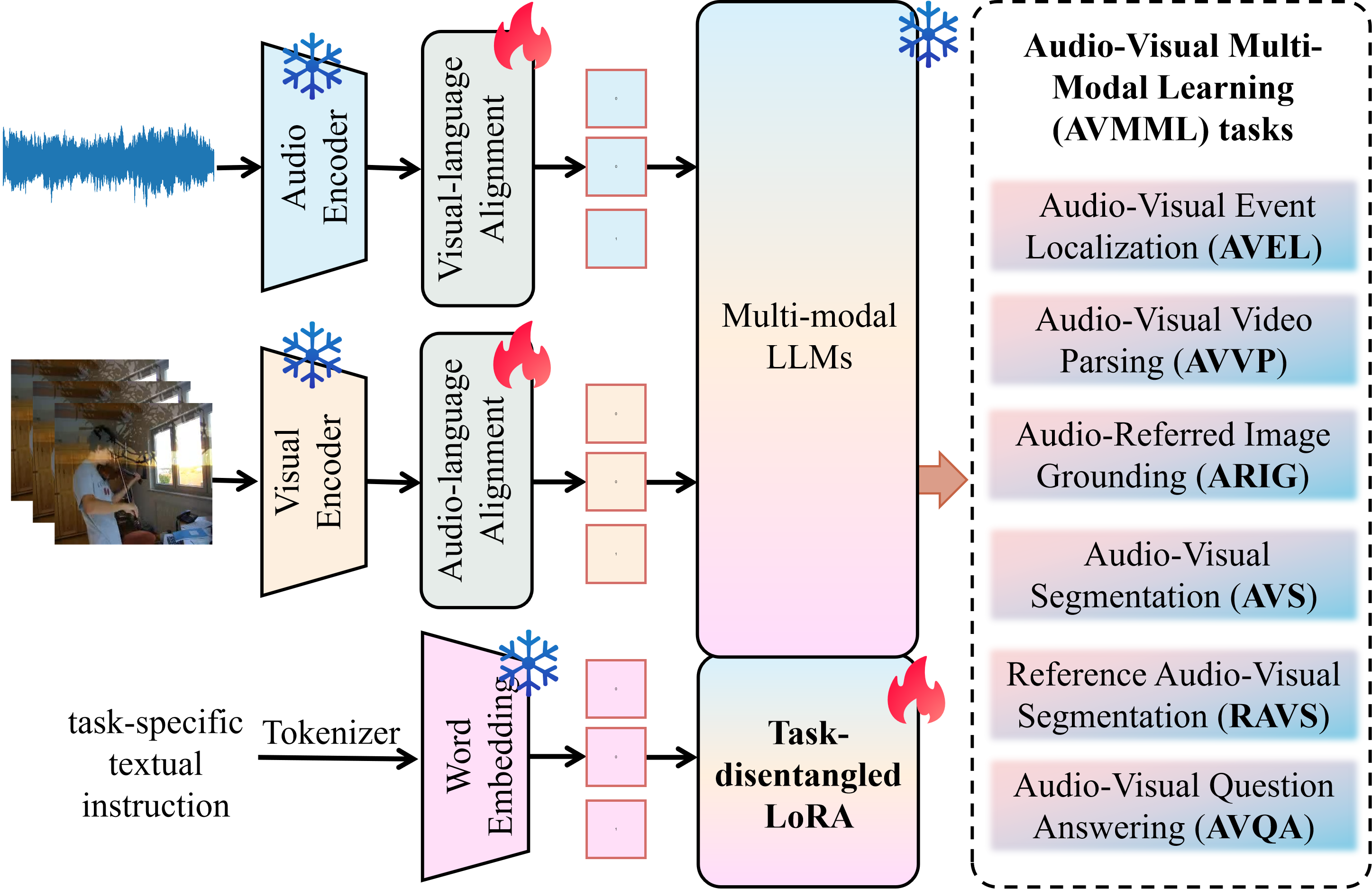}
  \caption{Overview of our proposed unified framework for versatile AVMML tasks.}
  \label{fig:overall_framework}
\end{figure}

\subsection{Preliminaries}

Consider an audio-visual video sequence $\{A_t,V_t\}_{t=1}^T$ 
partitioned into $T$ consecutive non-overlapping temporal segments,
where $A_t$ and $V_t$ denote the audio and visual components of the $t^{\text{th}}$ segment, respectively.
We adopt CLIP-ViT-L/14 \cite{radford2021learning} and BEATs \cite{chen2022beats}
as the visual and audio encoders, respectively, 
leveraging their strong pre-trained representations on large-scale vision and audio datasets, 
which have been widely validated for AVMML tasks. 
For the visual stream,
we extract patch-level embeddings for each temporal segment,
resulting in a sequence of segment-wise visual features
$\textit{\textbf{F}}_v = \{\textbf{\textit{v}}_i \in \mathbb{R}^{L_v \times D_v} \}_{i=1}^{T}$,
where $L_v$ and $D_v$ denote the number of visual tokens and token dimension, respectively.
Similarly,
we derive the segment-wise audio embeddings
$\textbf{\textit{F}}_a = \{\textbf{\textit{a}}_i \in \mathbb{R}^{L_a \times D_a} \}_{i=1}^{T}$,
with $L_a$ and $D_a$ representing  the number of audio tokens and the token dimension, respectively.

To bridge the cross-modality gap, 
we introduce two modality-specific alignment modules:
a visual-language alignment module and an audio-language alignment module.
Both modules project modality-specific features into a shared textual embedding space.
This enables the LLM to perform joint reasoning over both visual and audio cues within a unified representation space. 
Each alignment module comprises a Q-Former \cite{li2023blip} followed by an MLP projection layer.
Formally, the aligned visual and audio embeddings are computed as
\begin{equation}
\begin{aligned}
\textbf{\textit{H}}_v = \text{MLP}_v \bigl(\text{Q-Former}_v(\textbf{\textit{F}}_v)\bigr) \in \mathbb{R}^{T K_v \times h}, \\
\textbf{\textit{H}}_a = \text{MLP}_a \bigl(\text{Q-Former}_a(\textbf{\textit{F}}_a)\bigr) \in \mathbb{R}^{T K_a \times h},
\end{aligned}
\end{equation}
where $K_v$ and $K_a$ are the number of learnable query tokens for the visual and audio modalities, respectively,
and $h$ denotes the hidden dimension of the LLM embedding space.

For each AVMML task,
we formulate  task-specific textual instructions
by integrating a task description with its corresponding input query.
These instructions are tokenized and fed into the LLM's token embedding layer, 
yielding textual embeddings $\textbf{\textit{H}}_{txt} \in \mathbb{R}^{K_t \times h}$,
where $K_t$ denotes the number of text tokens.
The input $\textbf{\textit{H}}$ of multi-modal LLM
is then formed by replacing the audio-visual special tokens in the textual instruction with the aligned  visual and audio embeddings:
\begin{equation}
\begin{aligned}
\textbf{\textit{H}} = g(\textbf{\textit{H}}_v , \textbf{\textit{H}}_a , \textbf{\textit{H}}_{txt}) \in \mathbb{R}^{N \times h},
\label{Eq2}
\end{aligned}
\end{equation}
where $g(\cdot)$ represents the composite operation of special token replacement, insertion, and concatenation.
Here, $N=T K_v + T K_a+K_t$ denotes the total number of input tokens fed into the multi-modal LLM.

\subsection{Task-disentangled Low-Rank Adaptation}

\subsubsection{Motivation}
Although multi-modal LLMs 
% pre‑trained on large‑scale, general‑domain datasets 
exhibit strong generalization capability, 
adapting them to diverse AVMML tasks 
remains challenging
when task-specific training data is limited.
LoRA \cite{hu2022lora,yang2024low} 
exploits  the low intrinsic dimensionality of weight updates in LLMs. 
It decomposes the dense weight update matrix
$\Delta \textbf{\textit{W}} \in \mathbb{R}^{k \times h}$ into two low-rank matrices, 
$\textbf{\textit{B}} \in \mathbb{R}^{k \times r}$ and
$\textbf{\textit{A}} \in \mathbb{R}^{r \times h}$, 
with the rank constraint $r \ll \min(h, k)$. 
This decomposition significantly reduces the total number of trainable parameters during task-specific adaptation.
% The adapted weight matrix is then expressed as 
% $\textbf{\textit{W}}+\textbf{\textit{B}} \textbf{\textit{A}}$,
% where $\textbf{\textit{W}}$ is the original  pre-trained weight matrix.

However,
vanilla LoRA exhibits limited capacity to facilitate effective cross-task collaboration.
Inspired by HydraLoRA \cite{tian2024hydralora}, 
Du et al. \cite{du2025crab} design an asymmetric LoRA architecture 
comprising one shared  matrix $\textbf{\textit{A}}$ and 
$B$ independent LoRA heads $\{\textbf{\textit{B}}_i \in \mathbb{R}^{k \times r}\}_{i=1}^B$.
The weight update matrix is defined as
$\Delta \textbf{\textit{W}} = \sum_{i=1}^B u_i \textbf{\textit{B}}_i \textbf{\textit{A}}$,
where $\{u_i\}_{i=1}^B$ denote router scores, computed via a trainable projection layer followed by a softmax activation function.

Nevertheless,
this design \cite{du2025crab} still struggles to adequately capture the dynamic interplay between task-shared and task-specific knowledge across diverse AVMML tasks.
This limitation stems primarily from two underlying factors:
\begin{itemize}
\item Task-specific inputs are indiscriminately projected  into the identical low-dimensional subspace spanned by the matrix $\textbf{\textit{A}}$,
thereby undermining the discriminability of task-specific representations.
\item The routing mechanism determines  the activation of expert heads $\textbf{\textit{B}}_i$ solely based on the input $\textbf{\textit{H}}$,
which inevitably induces mutual interference across distinct  AVMML tasks.
\end{itemize}

\subsubsection{Architecture Design}
\begin{figure}[t]
  \centering
  \includegraphics[width=0.85\linewidth]{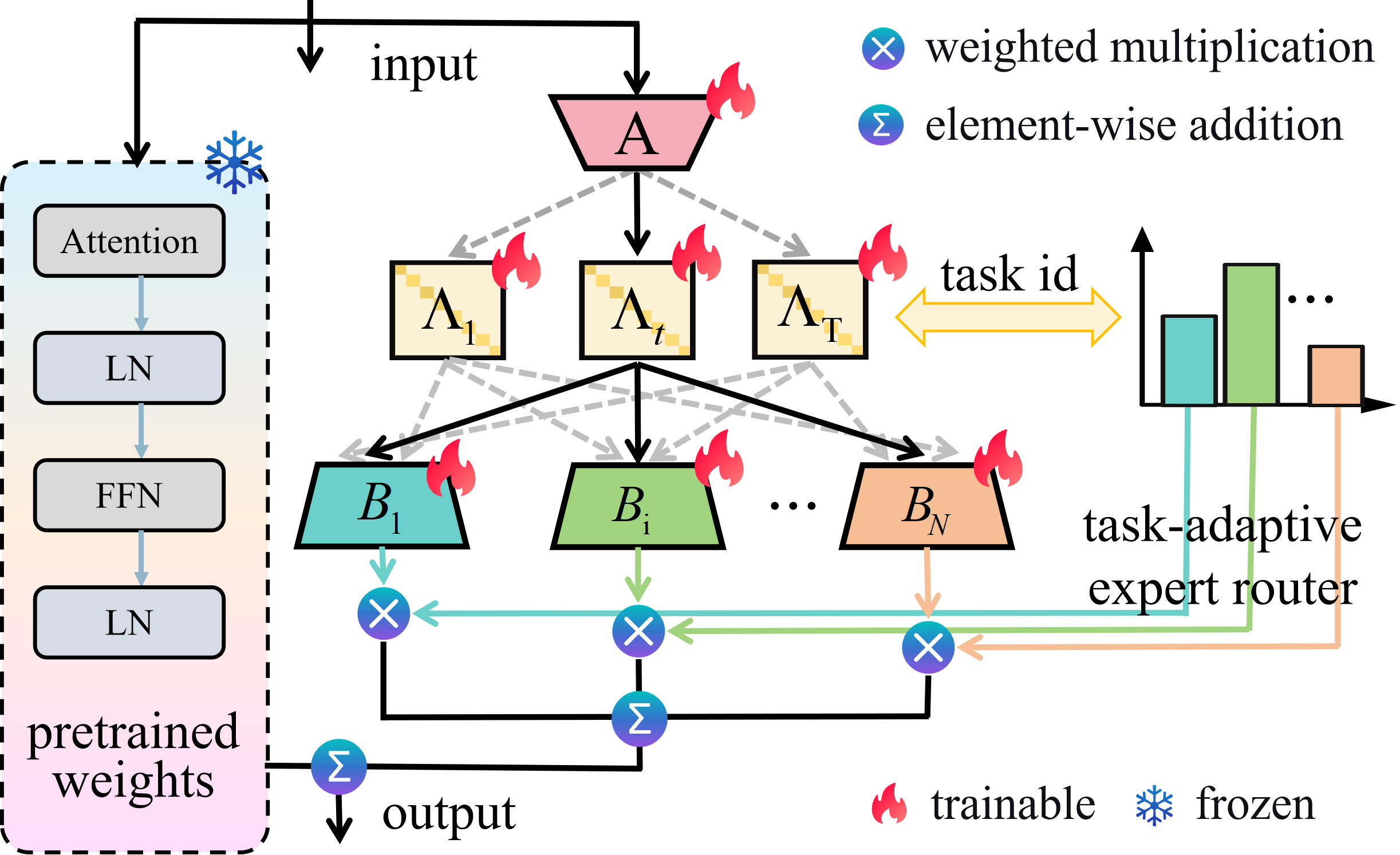}
  \caption{Architecture of the proposed task-disentangled LoRA,
  where $\textbf{\textit{A}}$,
  $\mathbf{\Lambda}_t$, and $\textbf{\textit{B}}_i$ denote task-general low-rank matrix, task-specific modulation matrices, and cross-task collaboration expert heads, respectively.
  % Here, $1 \leq t \leq T$ and $1 \leq i \leq N$.
  }
  \label{fig:MTL-LoRA module}
\end{figure}

To address this limitation, 
we propose a task-disentangled LoRA that enables dynamic adaptation and integration of task-shared and task-specific knowledge across diverse AVMML tasks,
as detailed in Fig.\ref{fig:MTL-LoRA module}.
Formally,
given the input for task $t$ 
derived from Eq.\ref{Eq2}, 
we reformulate it as $\textbf{\textit{H}}_t \in \mathbb{R}^{N \times h}$.
Similar to LoRA \cite{hu2022lora,yang2024low} and HydraLoRA \cite{tian2024hydralora},
our task-disentangled LoRA first projects $\textbf{\textit{H}}_t$ into a low-dimensional subspace 
through the task-general matrix $\textbf{\textit{A}} \in \mathbb{R}^{r \times h}$ 
to capture universal audio-visual knowledge.
To enhance task discriminability within this subspace
and preserve task-specific information,
we introduce $T$ task-specific modulation matrices
$\{\mathbf{\Lambda}_t \in \mathbb{R}^{r \times r}\}_{t=1}^T$,
which decouples task-specific patterns from the shared subspace.

Furthermore, we believe that adaptive information sharing is critical for effectively leveraging cross‑task transferable knowledge to boost overall task performance \cite{zhang2021survey,yang2025mtl}.
Accordingly,
inspired by the MoE mechanism \cite{pmlr-v162-du22c},
we introduce $B$ cross-task collaboration expert heads
$\{\textbf{\textit{B}}_i \in \mathbb{R}^{k \times r}\}_{i=1}^B$ to exploit inherent inter-task
correlations and enable cross-task
collaboration.
These experts
are adaptively aggregated via a task-adaptive weighted combination mechanism.
This design enables fine-grained adaptive information sharing across diverse AVMML tasks.
Formally,
let $\textbf{\textit{u}}_t \in \mathbb{R}^{n \times 1}$
denote the task-adaptive gating weight for task $t$.
The output of the proposed task-disentangled LoRA is formulated as:
\begin{equation} \label{eq:mtl_lora_forward}
\begin{aligned}
\textbf{\textit{H}}_t^{out} &= \left( \textbf{\textit{W}} + \Delta \textbf{\textit{W}}_t \right) \textbf{\textit{H}}_t \\
    &= \textbf{\textit{W}} \textbf{\textit{H}}_t 
    + 
    \sum_{i=1}^{B} 
    \frac{\exp(\textbf{\textit{u}}_t^i / \tau) 
    \textbf{\textit{B}}_i}
    {\sum_{j=1}^{n} \exp(\textbf{\textit{u}}_t^j / \tau)} 
    \mathbf{\Lambda}_t 
    \textbf{\textit{A}} 
    \textbf{\textit{H}}_t,
\end{aligned}
% \tag{4}
\end{equation}
where $\tau$ is a temperature parameter that controls the sharpness of the weight distribution.

\subsubsection{Analysis and Discussion}
We conduct a systematic analysis of the proposed  task-disentangled LoRA, focusing on its initialization strategy and gradient backpropagation. 
Complementing this analysis,
we present a geometric interpretation to elucidate the mechanism by which our task-disentangled LoRA facilitates task-specific feature separation.

\textbf{Initialization strategy.} 
Each task-specific low-rank matrix
$\mathbf{\Lambda}_t$ is initialized as an identity matrix. 
This strategy preserves the original matrix multiplication output at initialization, 
thus avoiding destabilizing perturbations induced by per-task parameterization in the early training phase.
As training progresses,
each $\mathbf{\Lambda}_t$ 
gradually deviates from its initial identity state,
enabling the model to capture discriminative task-specific patterns within the shared low-rank subspace.

\textbf{Task-aware gradient preservation.} 
In contrast to LoRA \cite{hu2022lora,yang2024low} and HydraLoRA \cite{tian2024hydralora},
which lack explicit task assignment mechanisms,
our task-disentangled LoRA introduces a task-specific matrix $\mathbf{\Lambda}_t$ as an independent learnable component for each task.
These independent components preserve distinct gradient descent directions for individual tasks throughout the optimization process,  
thereby enhancing robustness against inter-task interference.

\textbf{Geometric interpretation.} 
For ease of derivation,
consider an input $\textbf{\textit{H}}$ 
processed by two distinct tasks $i$ and $j$
through a LoRA expert head $\textbf{\textit{B}}$. 
In standard LoRA \cite{hu2022lora,yang2024low}, 
both tasks operate within an identical low-rank subspace derived from
$\textbf{\textit{A}}  \textbf{\textit{H}}$. 
In contrast, 
our task-disentangled  LoRA introduces two task-specific matrices, $\mathbf{\Lambda}_i$ and $\mathbf{\Lambda}_j$,
which project the shared subspace into two distinct task-specific subspaces,
denoted as $\mathbf{\Lambda}_i \textbf{\textit{A}}  \textbf{\textit{H}}$ and $\mathbf{\Lambda}_j \textbf{\textit{A}}  \textbf{\textit{H}}$, respectively.
The angular similarity between the two task-specific subspaces is bounded  by the Cauchy‑Schwarz inequality \cite{bhatia1995cauchy} as follows:
\begin{equation}
\bigl| (\textbf{\textit{B}}
\mathbf{\Lambda}_i 
\textbf{\textit{A}} 
\textbf{\textit{H}})^\top 
(\textbf{\textit{B}} \mathbf{\Lambda}_j 
\textbf{\textit{A}} 
\textbf{\textit{H}}) 
\bigr|
\le \|
\textbf{\textit{B}} \mathbf{\Lambda}_i 
\textbf{\textit{A}} \textbf{\textit{H}}\| 
\|\textbf{\textit{B}} 
\mathbf{\Lambda}_j 
\textbf{\textit{A}} 
\textbf{\textit{H}}\|.
\end{equation}
It follows that the cosine similarity between the two task-specific subspaces satisfies
\begin{equation}
\frac{\bigl| (\textbf{\textit{B}} \mathbf{\Lambda}_i \textbf{\textit{A}} \textbf{\textit{H}})^\top (\textbf{\textit{B}}
\mathbf{\Lambda}_j 
\textbf{\textit{A}} 
\textbf{\textit{H}}) \bigr|}
{\|\textbf{\textit{B}} 
\mathbf{\Lambda}_i 
\textbf{\textit{A}} 
\textbf{\textit{H}}\|\|\textbf{\textit{B}} 
\mathbf{\Lambda}_j 
\textbf{\textit{A}} 
\textbf{\textit{H}}\|}
\le \frac{\bigl| 
(\textbf{\textit{B}} 
\textbf{\textit{A}} 
\textbf{\textit{H}})^\top 
(\textbf{\textit{B}} 
\textbf{\textit{A}} 
\textbf{\textit{H}}) \bigr|}
{\|\textbf{\textit{B}} 
\textbf{\textit{A}} \textbf{\textit{H}}\|\|\textbf{\textit{B}} \textbf{\textit{A}} \textbf{\textit{H}}\|} = 1,
\label{Eq4}
\end{equation}
where the strict inequality holds if and only if  $\textbf{\textit{B}} \mathbf{\Lambda}_i \textbf{\textit{A}} \textbf{\textit{H}}$ is not collinear with $\textbf{\textit{B}} \mathbf{\Lambda}_j \textbf{\textit{A}} \textbf{\textit{H}}$.

When $\mathbf{\Lambda}_i$ and $\mathbf{\Lambda}_j$ converge to distinct transformations,
the cosine similarity is strictly smaller than the standard LoRA \cite{hu2022lora,yang2024low},
as derived  in Eq.\ref{Eq4}.
This observation suggests  that the task-specific modulation matrices introduce additional degrees of freedom to enhance cross-task feature discriminability, thereby alleviating inter-task interference.

\begin{table}[t]
\vspace{-0.3cm}
\centering
\caption{Dataset statistics for  the AVMML tasks considered in this paper. 
AVSBench comprises three subsets: 
Multiple Sound Source Segmentation (MS3) and
Single Sound Source Segmentation (S4) for the AVS, 
and ARIG for the ARIG.}
\vspace{-0.3cm}
\begin{threeparttable}
\setlength{\tabcolsep}{9pt} 
\begin{tabular}{ccccc}
\toprule
\multirow{2}{*}{\textbf{Dataset}} & \multirow{2}{*}{\textbf{Subset}} & \multirow{2}{*}{\textbf{Task}} & \multicolumn{2}{c}{\textbf{Num}.} \\ \cmidrule{4-5}
 &  &  & \textbf{train} & \textbf{test} \\
\midrule
AVE~\cite{tian2018audio}\tnote{1} & -- & AVEL & 3,339 & 402 \\
LLP~\cite{tian2020unified}\tnote{2} & -- & AVVP & 9,043 & 1,062 \\
MUSIC-AVQA~\cite{li2022learning}\tnote{3} & -- & AVQA & 31,923 & 9,129 \\
\multirow{3}{*}{AVSBench~\cite{zhou2022audio}\tnote{4}} & ARIG & ARIG & 3,452 & 3,694 \\
 & MS3 & \multirow{2}{*}{AVS} & 1,480 & 320 \\
 & S4 &  & 3,452 & 740 \\
Ref-AVS~\cite{wang2024ref}\tnote{5} & -- & RAVS & 14,113 & 1,656 \\
\midrule
\multicolumn{3}{c}{\textbf{Total}} & 66,802 & 17,003 \\
\bottomrule
\end{tabular}
\begin{tablenotes}[flushleft]
\footnotesize
\item[1] \url{https://github.com/YapengTian/AVE-ECCV18}
\item[2] \url{https://github.com/YapengTian/AVVP-ECCV20}
\item[3] \url{https://github.com/GeWu-Lab/MUSIC-AVQA}
\item[4] \url{https://github.com/OpenNLPLab/AVSBench}
\item[5] \url{https://github.com/GeWu-Lab/Ref-AVS}
\end{tablenotes}
\end{threeparttable}
\label{tab:I}
\vspace{-0.3cm}
\end{table}

\subsection{Segmentation Extension and Training}

Most existing  multi-modal LLMs
% such as CLIP \cite{radford2021learning}, BEATs \cite{chen2022beats} and LLaMAs \cite{touvron2023llama,cheng2024videollama}, 
are inherently  limited to generating textual outputs and lack the ability to produce segmentation masks.
Recent  works \cite{lai2024lisa,ren2024pixellm}
have addressed this limitation by integrating the mask decoder of the Segment Anything Model (SAM) \cite{kirillov2023segment} into multi-modal LLMs.
Inspired by these advances,
we introduce an ``\textit{embedding as mask}'' paradigm
to equip the multi-modal LLM with segmentation capabilities.
Specifically, 
we first extend the LLM vocabulary with a set of $N$ learnable mask tokens $\{\langle \texttt{MASK} \rangle_i\}_{i=0}^{N}$ that indicate segmentation outputs.
When the model generates a segmentation mask, 
its  output would include special  tokens.
We then extract the last-layer embeddings $\textbf{\textit{H}}_{mask}$ of the multi-modal LLM corresponding to these tokens. 
These embeddings serve as a prompt 
and are fed into the mask decoder  
$\mathcal{D}(\cdot)$ together with the visual embeddings $\textbf{\textit{F}}_v$.
The segmentation mask $\widehat{\textbf{\textit{M}}} \in \mathbb{R}^{T \times C \times H \times W}$ can be generated  as:
\begin{equation}
\widehat{\textbf{\textit{M}}}=\mathcal{D}(\textbf{\textit{H}}_{mask},\textbf{\textit{F}}_v),
\end{equation}
where $C$ represents the number of semantic categories, $H$ and $W$ represent height and width
of the input visual frame, respectively.

The overall training objective consists of 
an auto-regressive cross-entropy loss $\mathcal{L}_{txt}$ 
for text generation across all AVMML tasks,
and an auxiliary segmentation loss $\mathcal{L}_{seg}$
tailored for mask prediction.
The segmentation loss $\mathcal{L}_{seg}$ integrates 
a binary cross‑entropy term
$\mathcal{L}_{bce}$ 
and a Dice loss term $\mathcal{L}_{dice}$.
The complete training objective is formulated  as:
\begin{equation}
\begin{aligned}
\mathcal{L} &= \lambda_{txt} \cdot \mathcal{L}_{txt} + \lambda_{seg} \cdot \mathcal{L}_{seg},\\
\text{where} \quad
&\mathcal{L}_{seg} = \lambda_{bce} \cdot \mathcal{L}_{bce} + \lambda_{dice} \cdot \mathcal{L}_{dice}.
\end{aligned}
\label{Eq6}
\end{equation}
Here, $\lambda_{txt}$, $\lambda_{seg}$, $\lambda_{bce}$, and $\lambda_{dice}$ are balancing hyper-parameters.
\section{Experiments}

\subsection{Experimental Settings}

\subsubsection{Datasets}

Tab.\ref{tab:I} summarizes the statistics of all datasets used  for each task.
For AVEL,
we use the AVE dataset~\cite{tian2018audio}, which contains 28 event categories covering diverse real-life scenarios.
For AVVP,
we adopt the LLP dataset~\cite{tian2020unified}, 
a large-scale weakly-supervised benchmark comprising  real-world videos across 25 event categories.
For AVQA, 
the evaluation is conducted on MUSIC-AVQA dataset~\cite{li2022learning}, 
which comprises over 150 hours of real-world videos annotated with 9 question types and 33 distinct question templates.
For AVS, we employ the AVSBench dataset~\cite{zhou2022audio}, specifically its S4 and MS3 subsets, 
which cover 23 event categories,
including sounds emitted by humans, animals, vehicles, and musical instruments.
The S4 subset is designed for single-source scenarios, while MS3 targets multi-source cases.
For RAVS,
we leverage the Ref-AVS dataset~\cite{wang2024ref},
which spans 70 real-world event categories.
A notable  characteristic of Ref-AVS is that binary segmentation masks are augmented with semantic category information to produce semantic masks.
For ARIG,  we also draw from AVSBench~\cite{zhou2022audio}, using its ARIG subset where pixel-level masks of sounding objects are converted into bounding box annotations.
We adopt the AV-UIE dataset \cite{du2025crab} for instruction tuning.
This corpus is derived from the five datasets in Tab.\ref{tab:I} through explicit reasoning augmentation, where structured reasoning chains are generated via in-context learning with pre-trained multi-modal LLMs.
% During construction, consistency between the transformed and original labels is strictly maintained; low-quality samples are subsequently filtered out, and manual correction is applied for quality control. 
% The resulting reasoning chains carry rich spatio‑temporal cues, which substantially benefit downstream localization and reasoning tasks.

\begin{table*}[t]
\vspace{-0.3cm}
\centering
\caption{Comprehensive comparison with general-purpose methods across multiple  AVMML tasks. 
The symbol ~\cmark~ indicates that  the method is capable of the corresponding task,
though  no evaluation is provided in the original paper.
The symbol ~\xmark~ denotes that the method lacks such capability.
The dataset used for each specific task is detailed in Tab.\ref{tab:I}.
}
\vspace{-0.3cm}
\setlength{\tabcolsep}{8pt} 
\begin{tabular}{c|c|cc|cc|c|cc|cc|cc}
\toprule
\multirow{2}{*}{\textbf{Method}}
& \textbf{AVEL}
& \multicolumn{2}{c|}{\textbf{AVVP}}
& \multicolumn{2}{c|}{\textbf{ARIG}}
& \textbf{AVQA}
& \multicolumn{2}{c|}{\textbf{AVS (MS3)}}
& \multicolumn{2}{c|}{\textbf{AVS (S4)}}
& \multicolumn{2}{c}{\textbf{RAVS}} \\
& \textit{acc.}
& \textit{Seg.-level} & \textit{Event-level}
& \textit{cIoU} & \textit{AUC}
& \textit{acc.}
& \textit{mIoU} & \textit{F1}
& \textit{mIoU} & \textit{F1}
& \textit{mIoU} & \textit{F1} \\
\midrule

TimeChat \cite{ren2024timechat}
& \cmark & 51.3 & \cmark & \xmark & \xmark & \xmark
& \xmark & \xmark & \xmark & \xmark & \xmark & \xmark \\

MEERKAT \cite{chowdhury2024meerkat}
& \cmark & 55.0 & \cmark & \cmark & \cmark & \cmark
& \xmark & \xmark & \xmark & \xmark & \xmark & \xmark \\

GroundingGPT \cite{2024GroundingGPT}
& \cmark & \cmark & \cmark & 44.0 & 45.1 & \cmark
& \xmark & \xmark & \xmark & \xmark & \xmark & \xmark \\

X-InstructBLIP \cite{panagopoulou2023xinstructblip}
& \xmark & \xmark & \xmark & \xmark & \xmark & 44.5
& \xmark & \xmark & \xmark & \xmark & \xmark & \xmark \\

VALOR \cite{chen2024valor}
& \xmark & \xmark & \xmark & \xmark & \xmark & 78.9
& \xmark & \xmark & \xmark & \xmark & \xmark & \xmark \\

AnyRef \cite{he2024multi}
& \xmark & \xmark & \xmark & \xmark & \xmark & \xmark
& 55.6 & 60.3 & \cmark & \cmark & \cmark & \cmark \\

\midrule
Crab  \cite{du2025crab}
& 74.0 & 55.9 & {49.0} & 39.4 & {40.1} & \textbf{76.4}
& {58.2} & 66.3 & {73.3} & {86.8} & {45.6} & {63.0} \\

\textbf{Ours}
& \textbf{77.8} & \textbf{60.1} & \textbf{53.2}
& \textbf{41.1} & \textbf{41.7} & {76.1}
& \textbf{59.6} & \textbf{66.5}
& \textbf{74.7} & \textbf{87.8}
& \textbf{52.1} & \textbf{69.5} \\
\bottomrule

\end{tabular}
\label{tab:II}
\end{table*}

\begin{table*}[htbp]
% \scriptsize
\vspace{-0.3cm}
\centering
\caption{Performance Comparison under fully-supervised and weakly-supervised Settings (\%)}
\vspace{-0.3cm}
% \label{tab:fs_ws}
\setlength{\tabcolsep}{7.5pt} 
\begin{tabular}{cccccccccccc}
\toprule
\multirow{1}{*}{\textbf{Method}} & AVEL \cite{tian2018audio} &AVSDN \cite{linWang2019}  &AVT \cite{lin2020audiovisual}  &CMRAN \cite{xu2020cross}  &CAM \cite{Owens_2018_ECCV}
&PSP \cite{Zhou_2021_CVPR}  &MM-Pyramid \cite{yu2022mm}   &Crab \cite{du2025crab}  &\textbf{Ours} \\
\midrule
{ acc. (fully)} & 72.7 & 75.2  & 75.8 & 77.4 & 72.3 & \textbf{77.8} & \textbf{77.8} & 74.0  & \textbf{77.8}\\
{acc. (weakly)} & 66.7  & 69.4 & 70.2 & 73.0 & 68.8 & 73.5  & 73.2 & 70.3 & \textbf{74.0}\\
\bottomrule
\end{tabular}
\label{tab:III}
\end{table*}

% \subsubsection{Evaluation Metrics}
% We strictly adhere to the official evaluation protocols established for each AVMML task. 
% Classification accuracy (acc.) serves as the primary metric for the AVE \cite{tian2018audio} and MUSIC-AVQA \cite{li2022learning} datasets to quantify overall prediction performance.
% For the LLP dataset~\cite{tian2020unified},
% performance is measured by segment(seg.)-level  and event-level F1 scores, where a predicted event is considered correct only if its temporal IoU with the ground-truth event exceeds $0.5$. 
% For the AVSBench-ARIG \cite{zhou2022audio} dataset,
% we adopt consensus IoU (cIoU) and AUC for quantitative evaluation.
% For the Ref-AVS \cite{wang2024ref}, AVSBench-MS3, and AVSBench-M4 \cite{zhou2022audio} datasets,
% mean IoU (mIoU) and the maximum $F_\beta$ score are employed as core indicators. 
% Following standard practice, we fix $\beta^2=0.3$ and sweep all confidence thresholds of predicted probability masks to select the optimal $F_\beta$ value.

\begin{table*}[t]
\vspace{-0.3cm}
\centering
\caption{
Comparison results of task-specific methods on the AVVP task using the LLP dataset.
Both fine-grained Seg.-level and Event-level metrics are reported.
Here, $F_A$, $F_V$, and $F_{AV}$
denote the F1 scores of audio-only, visual-only, and audio-visual events, respectively.
The \textit{type} is the average score across the three event types, while the \textit{event} is the overall F1 score computed across all evaluated audio and visual events.}
\vspace{-0.3cm}
\setlength{\tabcolsep}{13.5pt}
\begin{tabular}{c ccccc ccccc}
\toprule
\multirow{2}{*}{\textbf{Method}} & \multicolumn{5}{c}{\textbf{Seg.-level}} & \multicolumn{5}{c}{\textbf{Event-level}} \\
\cmidrule(lr){2-6} \cmidrule(lr){7-11}
 & $F_A$ & $F_V$ & $F_{AV}$ & type & event & $F_A$ & $F_V$ & $F_{AV}$ & type & event \\
\midrule
PPL \cite{rachavarapu2024prototype} & \textbf{62.5} & 55.3 & 52.3 & 56.0 & 58.3 & \textbf{55.4} & 51.1 & 46.9 & 50.9 & 50.6 \\
CML \cite{lamba2021crossmodal} & 60.8 & 54.8 & 50.0 & 55.2 & 56.2 & 53.4 & 51.6 & 44.3 & 49.8 & 51.2 \\
CM-PIE \cite{chen2024cm} & 61.7 & 55.2 & 50.1 & 55.7 & 56.8 & 53.7 & 51.3 & 43.6 & 49.5 & 51.3 \\
A$_{\text{cross}}$+V$_{\text{self}}$ \cite{pasi2022modalityBias} & 60.5 & 54.9 & 50.5 & 55.3 & 56.5 & 51.9 & 51.2 & 44.3 & 49.1 & 48.9 \\
MTSM \cite{wu2022perceptive} & 62.1 & 54.4 & 50.7 & 55.7 & 56.7 & 52.7 & 50.9 & 44.6 & 49.4 & 49.3 \\
HAN \cite{tian2020unified} & 60.1 & 52.9 & 48.9 & 54.0 & 55.4 & 51.3 & 48.9 & 43.0 & 47.7 & 48.0 \\
MA \cite{wu2021heterogeneous} & 59.8 & 57.5 & \textbf{52.6} & 56.6 & 56.6 & 52.1 & 54.4 & 45.8 & 50.8 & 49.4 \\
MGN \cite{mo2022mmgn} & 60.8 & 55.4 & 50.4 & 55.5 & 57.2 & 51.1 & 52.4 & 44.4 & 49.3 & 49.1 \\
Crab \cite{du2025crab} & 50.2 & 67.6 & 51.0 & 56.3 & 55.9 & 44.5 & 64.7 & 47.0 & 52.0 & 49.0 \\
\textbf{Ours} & 55.3 & \textbf{68.3} & 51.1 & \textbf{58.2} & \textbf{60.1} & 49.6 & \textbf{66.1} & \textbf{47.7} & \textbf{54.5} & \textbf{53.2} \\
\bottomrule
\vspace{-0.3cm}
\end{tabular}
\label{tab:IV}
\end{table*}

% \begin{table}[htbp]
% \centering
% \caption{
% Comparison results of task-specific methods on the ARIG task using the AVSBench dataset, with cIoU and AUC as the evaluation metrics.}
% \label{tab:dataset2_perf}
% \begin{tabular}{ccc}
% \toprule
% \textbf{Method} & cIoU & AUC \\
% \midrule
% LVS \cite{chen2021localizing} & 23.7 & 25.0 \\
% EZ-VSL \cite{mo2022localizing}    & 26.4 & 29.0 \\
% SSPL \cite{song2022self}      & 26.5 & 28.6 \\
% SSL-TIE \cite{liu2022exploiting}   & 26.6 & 29.5 \\
% SLAVC \cite{mo2022SLAVC}     & 26.7 & 29.4 \\
% MarginNCE \cite{park2023marginnce} & 26.9 & 30.0 \\
% HearTheFlow \cite{fedorishin2022hear} & 27.1 & 30.5 \\
% FNAC \cite{sun2023learning}      & 27.2 & 31.0 \\
% Crab \cite{du2025crab}
% & 39.4 & 40.1  \\
% \textbf{Ours}
% & \textbf{41.1} & \textbf{41.7}  \\
% \bottomrule
% \end{tabular}
% \end{table}

\begin{table*}[htbp]
\centering
\vspace{-0.3cm}
% \scriptsize
\caption{
Comparison results of task-specific methods on the ARIG task using the AVSBench dataset, with cIoU and AUC as the metrics.}
\vspace{-0.3cm}
\setlength{\tabcolsep}{4pt} 
\begin{tabular}{ccccccccccc}
\toprule
\textbf{Method} &LVS \cite{chen2021localizing} &EZ-VSL \cite{mo2022localizing} &SSPL \cite{song2022self}  &SSL-TIE \cite{liu2022exploiting}  &SLAVC \cite{mo2022SLAVC}  
&MarginNCE \cite{park2023marginnce}  &HearTheFlow \cite{fedorishin2022hear}  &FNAC \cite{sun2023learning}  &Crab \cite{du2025crab}  &\textbf{Ours}\\
\midrule
cIoU & 23.7  & 26.4  & 26.5 & 26.6  & 26.7 & 26.9 & 27.1 & 27.2 & 39.4 & \textbf{41.1} \\
AUC & 25.0 & 29.0  & 28.6 & 29.5 & 29.4 & 30.0 & 30.5 & 31.0 & 40.1 & \textbf{41.7}  \\
\bottomrule
\end{tabular}
% \vspace{-0.3cm}
\label{tab:V}
\end{table*}

\subsubsection{Implementation Details}
Each video is uniformly sampled with 10 frames,
all resized to $224 \times 224$.
For audio, 
each raw waveform is first resampled to 16 kHz. 
We then extract  128-dimensional log-mel filterbank features using a 25 ms Povey window with a 10 ms frame shift. 
We utilize LLaMA-2-7B-Chat \cite{touvron2023llama} as our base model. 
The visual-language and audio-language alignment module
each consist of a Q-Former~\cite{li2023blip} with 32 learnable query tokens followed by an MLP projector.
The task-guided LoRA structure is employed in all linear
layers with a rank of 8.
We optimize the model with AdamW using a learning rate of $1 \times 10^{-4}$. 
% The scaling factor, and dropout rate are set to 16 and 0.05, respectively. 
The first fine-tuning stage runs for 5 epochs with a per-device batch size of 8 and a gradient accumulation step of 8 on four NVIDIA A40 GPUs. 
For the mask decoder,
we use two $\langle \texttt{MASK} \rangle$ token groups corresponding to two
scales of visual features from visual encoder, 
specifically from the 14th and second-to-last layers.
Each group has three tokens.
For the segmentation stage, we further train the segmentation branch for 30 epochs with a per-device batch size of 8. 
The four hyper-parameters defined in Eq.\ref{Eq6} are set to 1.0, 0.5, 1.0, 0.5, 
respectively.

% \subsection{Quantitative Comparisons}

\subsection{Quantitative Comparisons with General-purpose Methods}
As summarized in Tab.\ref{tab:II},
the proposed framework outperforms other general-purpose models, 
achieving both broader task coverage and superior performance across the six AVMML tasks.
Existing methods generally suffer from pronounced task-specific limitations.
For instance,
TimeChat \cite{ren2024timechat} and 
GroundingGPT \cite{2024GroundingGPT}
cover AVEL and AVVP but lack segmentation capabilities.
MEERKAT \cite{chowdhury2024meerkat} fails on all AVS and RAVS tasks.
AnyRef \cite{he2024multi} focuses solely on segmentation and lacks  temporal localization or spatio-temporal reasoning capabilities.
While Crab~\cite{du2025crab} exhibits comparatively stronger cross-task generalization, 
our method consistently surpasses it across eleven evaluation metrics.
The most substantial performance gap is observed on the RAVS task,
with gains of +6.59 mIoU and +6.46 F1 score.
% which underscores the proposed method’s advantage in reference segmentation.
Notable improvements are also attained on other tasks,
with +3.73 accuracy. on AVE, as well as +4.20 segment-level and +4.22 event-level  F1 scores on AVVP.

% Table~\ref{tab:main_results} compares TG-MTL-LoRA with Crab, our primary baseline, under the identical unified multi-task training protocol. Both methods share the same backbone, input-output format, task definitions, training data, and test splits. Thus, the performance difference largely reflects the effect of replacing the original adaptation module with the proposed task-conditioned LoRA. We also list representative task-specific methods as reference results to contextualize the performance on individual benchmarks, but they are not direct competitors.

% Across the seven tasks, TG-MTL-LoRA outperforms Crab on 11 of the 12 reported metrics. The improvements cover temporal localization, spatial localization, and pixel-level understanding, showing that the proposed adaptation module brings broad benefits under the unified audio-visual multi-task setting. Compared with the original shared adaptation strategy, TG-MTL-LoRA explicitly injects task identity into the low-rank update paths, enabling the model to better accommodate diverse output structures, supervision signals, and cross-modal reasoning demands. The only exception is AVQA, where TG-MTL-LoRA shows a slight decrease in overall accuracy. We further analyze this result using fine-grained question subtypes in Table~\ref{tab:avqa_subtype}.

\begin{table}[t]
\vspace{-0.3cm}
    \centering
    \caption{Comparison results of task-specific methods on the AVS task. Results are reported on the AVSBench dataset, covering both the S4 and MS3 subsets, with mean IoU (mIoU) and F1 score as the evaluation metrics.}
    \vspace{-0.3cm}
\setlength{\tabcolsep}{14pt} 
    \begin{tabular}{ccccc}
        \toprule
        \multirow{2}{*}{\textbf{Method}} & \multicolumn{2}{c}{\textbf{S4}} & \multicolumn{2}{c}{\textbf{MS3}} \\
        \cmidrule(lr){2-3} \cmidrule(lr){4-5}
        & \textit{mIoU} & \textit{F1} & \textit{mIoU} & \textit{F1} \\
        \midrule
        MSSL \cite{qian2020multiple}      & 44.9 & 66.3 & 26.1 & 36.3 \\
        SST \cite{duke2021sstvos}        & 66.3 & 80.1 & 42.6 & 57.2 \\
        iGAN \cite{Mao2021TransformerTS}      & 61.6 & 77.8 & 42.9 & 54.4 \\
        LGVT \cite{zhang2021learning}      & 74.9 & 87.3 & 40.7 & 59.3 \\
        TPAVI \cite{zhou2022audio}    & 72.8 & 84.8 & 47.9 & 57.8 \\
        BAVS \cite{liu2024bavs}      & \textbf{78.0} & 85.3 & 50.2 & 62.4 \\
        Crab~\cite{du2025crab}
        & 73.3 & 86.8 & 58.2 & 66.3 \\
        \textbf{Ours}         & 74.7 & \textbf{87.8} & \textbf{59.6} & \textbf{66.5} \\
        \bottomrule
    \end{tabular}
\label{tab:VI}
\vspace{-0.5cm}
\end{table}

\subsection{Quantitative Comparisons with Task-specific Methods}
We evaluate the efficacy of our proposed framework on six representative AVMML tasks. 
% including AVEL, AVVP, ARIG, AVQA, AVS, and RAVS.
% Collectively, 
These tasks span a comprehensive spectrum of AVMML fields, encompassing 
temporal localization tasks such as AVEL and AVVP,
spatial localization tasks exemplified by ARIG, 
spatio-temporal reasoning tasks represented by AVQA, 
and pixel-level understanding tasks comprising AVS and RAVS.

\begin{figure*}[t]
\vspace{-0.3cm}
    \centering
    \includegraphics[width=0.9\linewidth]{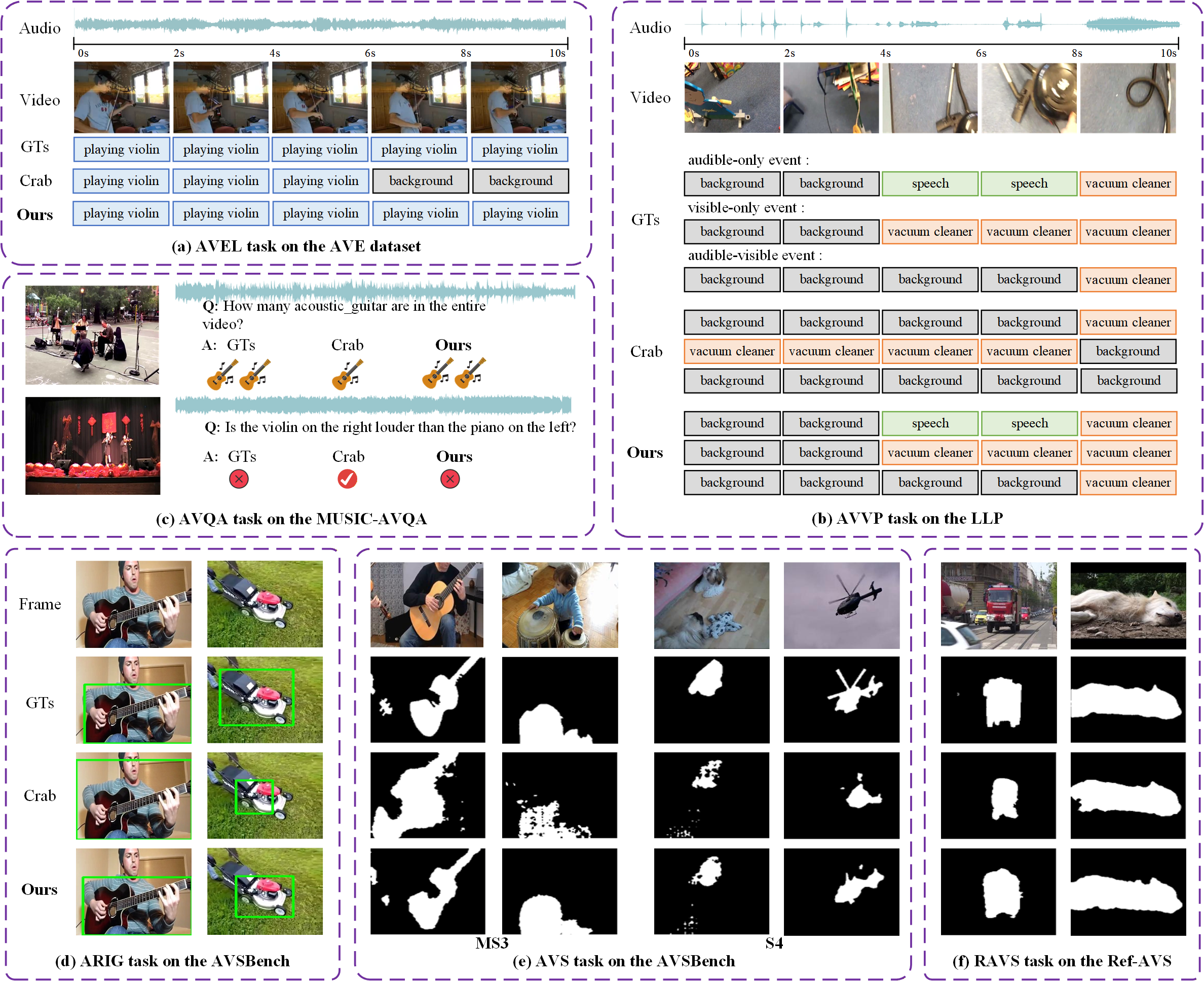}
\vspace{-0.35cm}
    \caption{Qualitative comparison of Crab \cite{du2025crab} and our method across six representative AVMML tasks.}
    \label{fig:spatial_qualitative}
    \vspace{-0.1cm}
\end{figure*}

\begin{table}[t]
\vspace{-0.3cm}
\centering
\caption{Comparison results of task-specific methods on the RAVS task. 
Results are reported on the Ref-AVS dataset, covering both the seen, unseen, and null subsets, with mean IoU (mIoU), F1 score, and null score (S) as the evaluation metrics.
Here, $\downarrow$ indicates lower values are  better.
Mix denotes the average over the Seen and Unseen subsets.}
\vspace{-0.3cm}
% \resizebox{\columnwidth}{!}{
\setlength{\tabcolsep}{4pt} 
\begin{tabular}{cccccccccc}
\toprule
\multirow{2}{*}{\textbf{Method}}
& \multicolumn{2}{c}{\textbf{Seen}}
& \multicolumn{2}{c}{\textbf{Unseen}}
& \multicolumn{2}{c}{\textbf{Mix}}
& \multicolumn{1}{c}{\textbf{Null}}\\
\cmidrule(lr){2-3} \cmidrule(lr){4-5} \cmidrule(lr){6-7}

& \textit{mIoU} & \textit{F1}
& \textit{mIoU} & \textit{F1}
& \textit{mIoU} 
& \textit{F1} 
& S$\downarrow$ \\

\midrule

AVSBench \cite{zhou2022audio}
& 23.2 & 51.1
& 32.4 & 54.7
& 27.8 & 52.9 
& 0.21 \\ 

R2VOS \cite{li2023robust}
& 25.0 & 41.0 
& 27.9 & 49.8
& 26.5 & 45.4  
& 0.18 \\ 

GAVS \cite{wang2024prompting}
& 28.9 & 49.8 
& 29.8 & 49.7 
& 29.4 & 49.8  
& 0.19  \\ 

ReferFormer \cite{wu2022language}
& 31.3 & 50.1
& 30.4 & 48.8 
& 30.9 & 49.5
& 0.17 \\ 

AVSegFormer \cite{gao2024avsegformer}
& 33.5 & 47.0
& 36.1 & 50.1
& 34.8 & 48.6 
& 0.17 \\

EEMC \cite{wang2024ref}
& 34.2 & 51.3 
& 49.5 & 64.8
& 41.9 & 58.1
& \textbf{0.01} \\ 

% VoCa \cite{guo2025leveraging}
% & 40.9 & \textbf{60.8} 
% & 51.7 & \textbf{70.8}
% & 46.3 & \textbf{66.8}
% & \textbf{0.005} \\ 

Crab~\cite{du2025crab}
& 40.5 & 58.0
& 45.6 & 63.0
& 43.1
& 60.5& \textbf{0.01}   \\
 
\textbf{Ours}
& \textbf{41.2} & \textbf{58.1}
& \textbf{52.1} & \textbf{69.5}
& \textbf{46.7} & \textbf{63.8} 
& \textbf{0.01}   \\
\bottomrule
\end{tabular}
% }
\label{tab:VII}
\vspace{-0.4cm}
\end{table}

\subsubsection{Temporal Localization}
Tab.\ref{tab:III} and Tab.\ref{tab:IV} 
present the comparative performance of the proposed framework  on two temporal localization tasks, i.e., AVEL and AVVP, respectively. 
On the AVEL task,
our method  achieves an accuracy of 77.8\%,
surpassing  previous  methods.
For the more challenging AVVP task,
our method attains the highest overall Type@AV and Event@AV scores at both segment and event levels, with leading $F_V$ scores underscoring its strength in visual event localization. 
Although PPL \cite{rachavarapu2024prototype} yields higher $F_A$ scores, 
the proposed method delivers balanced cross-modal performance and dominant comprehensive results while preserving  cross-task generalizability.

\subsubsection{Spatial Localization}
Tab.\ref{tab:V} presents the comparison on the ARIG task over the AVSBench dataset.
Task-specific methods exhibit limited overall performance, with cIoU ranging from 23.7 to 27.2 and AUC from 25.0 to 31.0.
Crab \cite{du2025crab} significantly outperforms these methods,
reaching 39.4 cIoU and 40.1 AUC.
The proposed method further improves to 41.1 cIoU and 41.7 AUC, surpassing Crab \cite{du2025crab} by +1.7 and +1.6, respectively.
This demonstrates the effectiveness of our framework in enhancing cross-modal alignment and spatial grounding.

\subsubsection{Pixel-level Understanding}
Tab.\ref{tab:VI} and Tab.\ref{tab:VII}
compare the proposed method with task-specific approaches on the AVS and RAVS tasks, respectively.
As shown in Tab.\ref{tab:VI},
while BAVS \cite{liu2024bavs}
tops the S4 subset in mIoU,
it underperforms on the more complex MS3 scenario.
By contrast,
our method delivers well-balanced segmentation performance across both single- and multi-source scenarios.
As presented  in Tab.\ref{tab:VII},
the most substantial improvement appears on the unseen subset, 
where  mIoU rises  from 45.6\% to 52.1\% and F1 from 63.0\% to 69.5\%, reflecting strong generalization to novel concepts.

\begin{table*}[t]
\vspace{-0.3cm}
\centering
\caption{Comparison results of task-specific methods on the AVQA task using the MUSIC-AVQA dataset. 
Fine-grained accuracy is reported under audio, visual,and audio-visual question modalities. 
Here, \textit{avg.} denotes the overall average accuracy across all question types,
including existential (\textit{exist.}), counting (\textit{count.}), localization (\textit{local.}), comparative (\textit{comp.}), and temporal (\textit{temp.}).}
\vspace{-0.3cm}
\setlength{\tabcolsep}{8pt}
\begin{tabular}{cccccccccccccc}

\toprule

\multirow{2}{*}{\textbf{Method}}
& \multicolumn{3}{c}{\textbf{Audio}}
& \multicolumn{3}{c}{\textbf{Visual}}
& \multicolumn{6}{c}{\textbf{Audio-Visual}} 
& \multirow{2}{*}{\textbf{avg.}}\\
\cmidrule(lr){2-4}\cmidrule(lr){5-7}\cmidrule(lr){8-13}

& \textit{count.} & \textit{comp.} &\textit{avg.} 
& \textit{count.}& \textit{local.} &\textit{avg.} 
& \textit{exist.} & \textit{count.}& \textit{local.} 
& \textit{comp.} & \textit{temp.} &\textit{avg.} 
& \\
\midrule

FCNLSTM \cite{2020Temporal}
& 70.8 & 65.7 & 68.9
& 64.6 & 48.1 & 56.2
& 82.3 & 60.0 & 46.2
& 62.9 & 47.5 & 60.4 & 60.8 \\

HME \cite{fan2019heterogeneous}
& 73.7 & 63.7 & 69.9
& 67.4 & 70.2 & 68.8
& 80.9 & 63.6 & 54.9
& 63.0 & 60.6 & 64.8 & 66.8 \\

PanoAVQA \cite{yun2021panoavqa}
& 75.7 & 66.0 & 72.1 
& 70.5 & 75.8 & 73.2
& 82.1 & 65.4 & 61.3
& 63.7 & 62.0 & 67.0 & 69.5 \\

COCA \cite{lao2023coca}
& 79.9 & \underline{67.7} & 75.4 
& 75.1 & 75.4 & 75.2
& \textbf{83.5} & 66.6 & 69.7
& \underline{64.1} & 65.6 & 70.0 & 72.3 \\

PSTP-Net \cite{li2023progressive}
& 74.0 & 65.6 & 70.9 
& 77.2 & 77.4 & 77.3
& 76.2 & 72.2 & \underline{71.8}
& 71.8 & \textbf{69.0} & \textbf{72.6} & 73.5 \\

LAVISH \cite{lin2023vision}
& 82.1 & 65.6 & \underline{76.0 }
& 79.0 & 81.4 & 80.2
& 81.7 & 75.5 & 66.1
& 63.8 & \underline{68.0} & 71.3 & 74.5 \\

APL \cite{li2024object} 
& 82.4 & \textbf{70.7} & \textbf{78.1} 
& 76.5 & 82.7 & 79.7
& \underline{83.0} & 66.9 & \textbf{73.3 }
& \textbf{64.8} & 66.0 & 71.0 & 74.5 \\

Crab \cite{du2025crab}
& \textbf{86.1} & 55.5 & 74.5 
& \underline{86.2} & \textbf{90.9} & \textbf{87.7}
& 82.7 & \textbf{80.7} & 68.6 
& 59.8 & 64.6 & \underline{71.7} & \textbf{76.4} \\

% \hline
\textbf{Ours}
& \underline{85.3} & 56.1 & 73.7
& \textbf{86.3}& \underline{90.0}  & \underline{87.5}
& 82.7 & \underline{78.9} & 68.3 
& 61.1 & 63.4 & 71.5 & \underline{76.1}\\

\bottomrule
\end{tabular}
\label{tab:VIII}
\vspace{-0.2cm}
\end{table*}

\begin{figure}[t]
    \centering
    \includegraphics[width=1.0\linewidth]{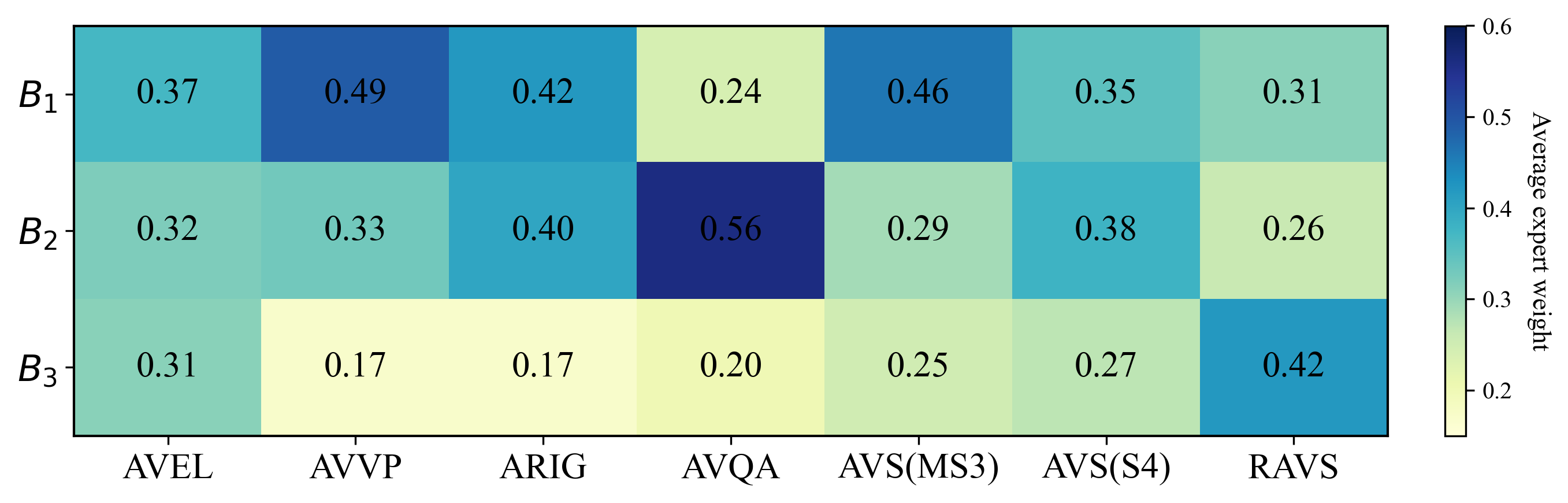}
    \vspace{-0.7cm}
    \caption{Visualization of task-specific expert preferences across diverse AVMML tasks. The heatmap shows the softmax-normalized mixture weights allocated  by each task to three LoRA heads in the \texttt{o\_proj} layer of the last decoder block. Darker colors denote stronger reliance of the corresponding task on the associated expert head.}
    \label{fig:headmap}
\end{figure}

\subsubsection{Spatio-temporal Reasoning}
Tab.\ref{tab:VIII} provides a fine-grained AVQA performance comparison on the MUSIC-AVQA dataset, 
broken down by modality and five question types.
Our method reaches 76.11\% overall accuracy, highly comparable to Crab \cite{du2025crab} and substantially superior to all task-specific approaches. 
It delivers the best visual counting performance, and posts modest improvements over Crab \cite{du2025crab} on comparative questions.

\subsection{Qualitative Results}

Fig.\ref{fig:spatial_qualitative} presents qualitative comparisons between  the proposed framework and the Crab \cite{du2025crab} across six representative AVMML tasks. 
For AVEL, Crab terminates events prematurely, while our method yields more accurate temporal boundaries aligned with GTs.
For AVVP, Crab suffers from high audio false positives and fails at visible-only event detection, 
whereas our method achieves accurate cross-modal event parsing.
For AVQA, our method correctly counts  acoustic guitars and answers comparative loudness questions accurately, 
demonstrating stronger quantitative and relational reasoning capabilities.
For ARIG, 
our method generates bounding boxes with higher GT overlap and more precise localization than Crab.
For both the AVS and RAVS tasks, 
our method produces cleaner and more complete segmentation masks, 
whereas Crab suffers from fragmented, noisy results with missing target regions.
These qualitative results verify the superiority of our framework in temporal localization, spatial grounding, pixel-level understanding, and spatio-temporal reasoning.

\begin{table}[t]
\vspace{-0.3cm}
\centering
\caption{Overall performance of the proposed method across all AVMML tasks under different evaluation settings.}
\vspace{-0.3cm}
\setlength{\tabcolsep}{13pt} 
\begin{tabular}{ccccc}
\toprule
\multirow{2}{*}{\textbf{Task}} &  & \multirow{2}{*}{\textbf{Metric}} & \multicolumn{2}{c}{\textbf{Expert Num.}}  \\
 &  &  & \textbf{four} & \textbf{{three}} \\
\midrule
\multirow{2}{*}{AVEL}
& fully & acc.  & 76.0 & \textbf{77.8} \\
&    weakly & acc. & 73.5 & \textbf{74.0} \\
% \midrule
\multirow{4}{*}{AVVP}
& \multirow{2}{*}{Seg.-level}  & type  & 58.0 & \textbf{58.2} \\
&                             & event & 60.0 & \textbf{60.1} \\
& \multirow{2}{*}{Event-level}& type  & 54.2 & \textbf{54.5} \\
&                             & event & 53.1 & \textbf{53.2} \\
% \midrule
\multirow{2}{*}{ARIG}
& \multirow{2}{*}{--} & cIoU & 40.3 & \textbf{41.1} \\
&                      & AUC  & 41.1 & \textbf{41.7} \\
% \midrule
AVQA & -- & acc. & \textbf{76.4} & 76.1 \\
% \midrule
\multirow{4}{*}{AVS}
& \multirow{2}{*}{MS3} & mIoU & \textbf{60.8} & 59.6 \\
&                       & F1   & \textbf{68.4} & 66.5 \\
& \multirow{2}{*}{S4}  & mIoU & 74.1 & \textbf{74.7} \\
&                       & F1   & 87.3 & \textbf{87.8} \\
% \midrule
\multirow{2}{*}{RAVS}
& \multirow{2}{*}{--} & mIoU & 39.0 & \textbf{46.7} \\
&                      & F1   & 53.9 & \textbf{63.8} \\
\bottomrule
\end{tabular}
\label{tab:IX}
\end{table}

\subsection{Mechanism Analysis}
To further understand how our method performs task-disentangled adaptation, 
we visualize the learned task-adaptive expert combination weights across all target tasks on their respective datasets in Fig.\ref{fig:headmap}. 
% In our task-guided LoRA,
% cross-task adaptation is realized 
% via task-specific aggregation of multiple low-rank expert pathways, 
% where the mixture weights inherently encode how each task selects and combines adaptation routes. 
From this visualization,
we can examine whether diverse AVMML tasks develop distinct update paths rather than converging to a uniform shared pattern. 
It also provides an intuitive characterization of task-specific expert preferences.

% We extract the expert mixture weights from the \texttt{o-proj} layer of the final decoder block, with the resulting heatmap shown in 
Fig.\ref{fig:headmap} clearly reveals  distinct  preference patterns across the three expert heads. 
Specifically, 
AVE, AVVP, and AVS (MS3) allocate relatively higher weights to $\textbf{\textit{B}}_1$,
while AVQA predominantly relies  on $\textbf{\textit{B}}_2$.
In contrast,
RAVS assigns the greatest weight to $\textbf{\textit{B}}_3$. 
ARIG distributes its weights primarily across both $\textbf{\textit{B}}_1$ and $\textbf{\textit{B}}_2$. 
These findings demonstrate that our task-disentangled LoRA effectively decouples task-specific adaptation, thereby enabling distinct update pathways for diverse AVMML tasks.
% our method learns task-adaptive expert combinations, 
% enabling distinct update pathways for diverse tasks.
% within a single unified framework.

\subsection{Ablation Studies}
\label{sec:ablation}

\subsubsection{Expert Number}

Tab.\ref{tab:IX} compares our three-expert method with the four-expert variant across six AVMML tasks.
Our method yields consistent gains on AVEL, AVVP and ARIG. 
The four-expert variant holds a marginal advantage on AVQA and AVS (MS3), 
while our method outperforms on AVS (S4) and delivers substantial improvements on RAVS.
Overall, the three-expert design achieves superior comprehensive performance on most tasks, striking an optimal balance between adaptation capacity and parameter efficiency.

% This is consistent with the ablation results in Section~\ref{sec:ablation}, where removing any expert head leads to performance degradation on multiple tasks and removing \(\Lambda_t\) causes consistent drops across all metrics. Together, the visualization and ablation results show that shared matrices \(A\) and \(B_i\) alone are insufficient to distinguish heterogeneous tasks, while the task-specific modulation matrix \(\Lambda_t\) enables more discriminative low-rank update paths under a largely shared parameter space.

\subsubsection{Task-specific Modulation matrix}
As shown in the first and last rows of Tab.\ref{tab:X},
removing the task-specific modulation matrix $\mathbf{\Lambda}$ degrades performance across all tasks, with the sharpest drops on temporal localization tasks AVEL and AVVP. 
ARIG and AVQA  also suffer measurable declines.
This confirms the necessity of 
$\mathbf{\Lambda}$, which injects task-specific transformations into shared expert heads to enable differentiated adaptation for diverse tasks.

\subsubsection{Expert Removal}
As shown in the last four rows of Tab.\ref{tab:X},
removing any expert consistently degrades performance degradation across all target tasks,
with the drop being particularly severe
when $\textbf{\textit{B}}_1$
is removed.
These results confirm that each expert contributes an irreplaceable, task-specific adaptation pathway.

\begin{table*}[t]
\vspace{-0.3cm}
\centering
\caption{Component ablation on the task-specific modulation matrix and LoRA expert heads.
The columns $\Lambda$, $B_1$, $B_2$, and $B_3$ denote the task-specific modulation matrix and the three LoRA expert heads, respectively.
``\checkmark'' indicates that the corresponding component is retained, whereas ``\xmark'' indicates that its corresponding weight is set to zero during inference process.}
\vspace{-0.3cm}
% \renewcommand{\arraystretch}{1.08}
% \setlength{\tabcolsep}{4.2pt}
% \resizebox{\linewidth}{!}{
\setlength{\tabcolsep}{12pt}
\begin{tabular}{cccc cc cccc cc c}

\toprule

\multicolumn{4}{c}{\textbf{Task}} 
& \multicolumn{2}{c}{\textbf{AVEL}} 
& \multicolumn{4}{c}{\textbf{AVVP}} 
& \multicolumn{2}{c}{\textbf{ARIG}} 
& \textbf{AVQA} \\

\cmidrule(lr){5-6}
\cmidrule(lr){7-10}
\cmidrule(lr){11-12}
\cmidrule(lr){13-13}

\multicolumn{4}{c}{\textbf{Module}}
& \textbf{fully} & \textbf{weakly}
& \multicolumn{2}{c}{\textbf{Seg.-level}}
& \multicolumn{2}{c}{\textbf{Event-level}}
& \multirow{2}{*}{\textit{cIoU}} & \multirow{2}{*}{\textit{AUC}}
& \multirow{2}{*}{\textit{acc.}} \\

\cmidrule(lr){1-4}
\cmidrule(lr){5-6}
\cmidrule(lr){7-8}
\cmidrule(lr){9-10}

$\mathbf{\Lambda}$ & $\textbf{\textit{B}}_1$ & $\textbf{\textit{B}}_2$ & $\textbf{\textit{B}}_3$
& \textit{acc.} & \textit{acc.}
& \textit{type} & \textit{event}
& \textit{type} & \textit{event}
&  &  &  \\
\midrule

\xmark & \cmark & \cmark & \cmark
& 75.0 & 71.6
& 57.4 & 59.2 & 53.1 & 51.8
& 40.9 & 41.5
& 75.3  \\

\cmark & \xmark & \cmark & \cmark
& 64.7 & 61.4
& 54.8 & 56.9 & 51.5 & 49.9
& 37.1 & 37.8
& 73.6  \\

\cmark & \cmark & \xmark & \cmark
& 67.1 & 63.4
& 54.3 & 55.8 & 50.9 & 49.2
& 36.3 & 37.1
& 73.1 \\

\cmark & \cmark & \cmark & \xmark
& 67.2 & 63.5
& 55.0 & 56.8 & 51.3 & 50.0
& 37.1 & 37.9
& 73.1  \\

\cmark & \cmark & \cmark & \cmark
& \textbf{77.8} & \textbf{74.0}
& \textbf{58.2} & \textbf{60.1} & \textbf{54.5} & \textbf{53.2}
& \textbf{41.1} & \textbf{41.7}
& \textbf{76.1} \\

\bottomrule
\end{tabular}
 % }
\label{tab:X}
\end{table*}

\section{Conclusion}
We present a unified versatile audio-visual multi-modal learning framework with a task-disentangled low-rank adaptation mechanism, which dynamically decouples task-specific and task-shared knowledge  to facilitate  effective multi-task collaboration. 
Experiments on six diverse tasks demonstrate that our method outperforms existing unified models and many task-specific approaches,
validating its superior performance, strong generalization, and universal effectiveness.

\bibliographystyle{IEEEtran}
\bibliography{reference}

\end{document}